# Gravitational Instability of Perturbations in a Background Nonlinear Nonstationary Model of a Disk-Like System.

## IV. Formation of two-ring structures in galaxies

K.T. Mirtadjieva[1,2], K.A. Mannapova[2]

[1] –Astronomical Institute of the Academy of Sciences of Uzbekistan, Astronomicheskaya 33, Tashkent 100052, Uzbekistan

[2]- National University of Uzbekistan, Physical Faculty Vuzgorodok, Tashkent 100174, Uzbekistan

**Abstract.** In order to identify the exact criteria for the formation of two-ring structures in galaxies, we studied the issue of gravitational instability of the corresponding structural vibration modes against the background of a composite disk model [1-3], which has an exact nonlinear law of nonstationarity. Nonstationary analogs of dispersion equations for the given structural vibration modes of the model are derived and the results of their analysis are obtained. A comparative analysis of the increments of instabilities of ring-like oscillation modes is carried out to determine the dependence of their characteristic times of manifestation on the main physical parameters of the model.

## 1. INTRODUCTION

Ring-like structures are observed in many types of astrophysical objects, ranging from planets [4] to galaxies (see [5] and refs. there). This suggests that the origin of these structures is associated, in particular, with certain general physical mechanisms, for example, with the evolution of natural oscillation modes. In [5], we developed a classification of ring-like galaxies to study their nature and diversity. A physical explanation of the features of ring-shaped galaxies and questions of their origin requires the construction of theoretical models and analysis of the gravitational instabilities of these structural vibration modes of the latter.

Assuming that global structural formations of galaxies can begin to form at the early non-stationary stage of their evolution, in we considered ring-like instability against the background of nonlinear models of a nonequilibrium self-gravitating disk [2,5]. These models are a nonlinear nonequilibrium generalization of the stationary model of Bisnovaty-Kogan and Zel'dovich [7] for the case of radial pulsations. The criteria for the formation of single-ring galaxies against the background of a composite nonlinear non-stationary model were studied by us in the previous part of this work [2] by studying the oscillation mode with the main harmonic index N = 4 and azimuthal wave number m = 0 and m = 2. But in our classification of ring-like galaxies [5], the two-ring galaxies accounted for rather more percent than expected. For this reason in this paper we investigated the origin of two-ring galaxies together with the study of the problem of gravitational instability of the corresponding structural oscillation modes on the background of a composite disk model [1-3].

Two-ring formations against the background of this model can form as a result of gravitational instability of the vibration mode with N = 6 and m = 0; 2. If at N = 6 and m = 0 we have purely ring structures, then at N = 6 and m = 2 the rings are split into separate condensations. Using the obtained results of the study, critical diagrams of the dependence of the initial virial ratio on the parameters of the composite model were constructed. Also, a comparative analysis of the increments of instabilities of the ring-shaped vibration modes is carried out to determine the dependence of their characteristic times of manifestation on the main physical parameters of the model.

## 2. BASIC RELATIONS AND EQUATIONS

In this part of the work, like the previous ones [1-3], we investigate the problem of gravitational instability of the following non-stationary model in the phase description with an anisotropic velocity diagram:

$$\Psi(\vec{r},\vec{v},\Omega,\lambda,\nu,t)=(1-\nu)\Psi_1(\vec{r},\vec{v},\lambda,\Omega,t)+\nu\Psi_2(\vec{r},\vec{v},\lambda,\Omega,t), \quad (1)$$

which enables us to investigate intermediate states between two different models covering broader possible initial conditions at an early non-stationary stage of evolution disco-shaped self-gravitating systems. Here $\nu$ is the superposition parameter, $\Omega$ – dimensionless parameter characterizing the value of the solid-state rotation of the disk, and the amplitude of the pulsation $\lambda=1-(2T/|U|)_0$ exactly expressed in terms of the values of the virial ratio at the moment in time t=0.

In the composite model (1) as $\Psi_1$ и $\Psi_2$ we took nonlinearly pulsating isotropic and anisotropic disk models [1-3,6]

$$\Psi_1=\frac{\sigma_0}{2\pi\Pi\sqrt{1-\Omega^2}}\left[\frac{1-\Omega^2}{\Pi^2}\left(1-\frac{r^2}{\Pi^2}\right)-(v_r-v_a)^2-(v_\perp-v_b)^2\right]^{-1/2}, \quad (2)$$

$$\Psi_2=\frac{\sigma_0}{\pi}\left[1+\Omega\cdot r\cdot v_\perp)\right]\cdot\chi(D) \quad , \quad (3)$$

where $\sigma_0$- the value of the surface density of the disk at t = 0, r = 0, $\Pi(t)=(1+\lambda\cos\psi)\cdot\left(1-\lambda^2\right)^{-1}$ - the stretching factor of the system, and $\psi$ is an auxiliary variable connected to time t as follows: $t=(\psi+\lambda\sin\psi)\cdot\left(1-\lambda^2\right)^{-3/2}$ , $\chi$ is a Heaviside function. Normalization is accepted everywhere $\pi^2 G\sigma_0=2R_0$ ($R_0$=1), and the quantities $\lambda$ and $\Omega$, like $\nu$, take values from the interval [0; 1], $v_r$ and $v_\perp$ – radial and tangential components of the "particle" velocity with the coordinate $\vec{r}(x, y)$, the quantity D is $D=\left(1-r^2/\Pi^2\right)\left(1-\Pi^2 v_\perp^2\right)-\Pi^2(v_r-v_a)^2$ , and

$$v_a=-\lambda\frac{r\sin\psi}{\sqrt{1-\lambda^2\Pi^2}}, \qquad v_b=\frac{\Omega r}{\Pi^2} \quad . \tag{4}$$

The composite model (1) has the following surface density

$$\sigma(\vec{r}, t)=\frac{\sigma_0}{\Pi^2(t)}\sqrt{1-\frac{r^2}{\Pi^2(t)}} \tag{5}$$

and performs radial pulsations with the period

$$P(\lambda)=\frac{2\pi}{\left(1-\lambda^2\right)^{3/2}} \quad . \tag{6}$$

Let us note that by analogy with the theory of stability of equilibrium models, to analyze and find criteria for the instability of a nonlinearly non-equilibrium model, it is necessary to derive a nonstationary analogue of the dispersion equation (NADE). And to obtain a composite model (1) corresponding to NADE, a small asymmetric perturbation with a potential $\delta\Phi$ is superimposed on it, and taking this into account, in [1,2,6] we have given the basic equation for the centroid displacement vector: $\overline{\delta\vec{r}}$ :

$$\Lambda\overline{\delta\vec{r}}=\left[(1+\lambda\cos\psi)\frac{d^2}{d\psi^2}+\lambda\sin\psi\frac{d}{d\psi}+1\right]\overline{\delta\vec{r}}=\Pi^3(\psi)\overline{\frac{\partial(\delta\Phi)}{\partial\vec{r}}} \quad , \tag{7}$$

where the bar above denotes averaging over the velocity space. The solution to equation (7) can be represented in integral form [1,2,6]

$$\overline{\delta\vec{r}}=\int_{-\infty}^{\psi}\Pi^3(\psi_1)S(\psi,\psi_1)\overline{\left[\frac{\partial(\delta\Phi)}{\partial\vec{r}}\right]}d\psi_1 \ , \tag{8}$$

moreover, $S(\psi, \psi_1)$ is an analogue of the Green's function, which is constructed in the standard way from the solution of the homogeneous equation in (7) and is equal to

$$S(\psi, \psi_1)=\left[\sin\psi(\cos\psi_1+\lambda)-\sin\psi_1(\cos\psi+\lambda)\right]\left(1+\lambda\cos\psi_1\right)^{-2} . \tag{9}$$

Now it needs to clarify the form of the perturbation $\delta\Phi$. Note that ring-like modes belong to the class of horizontal perturbations that develop only in the plane of the disk (x, y) and do not depend on z. Taking into account the nature of the investigated non-stationary model (1), by analogy with the theory of stability of stationary models [8, 9], these oscillations can be described in the form

$$\delta\Phi = D_{mN}(\psi)\ r^{N-m}(x+iy)^m \qquad (r=\sqrt{x^2+y^2})\ , \tag{10}$$

where $D_{mN}(\psi)$– is the required function, which, in contrast to the case of stationary models, depends on time.

Now, to derive NADE, it is necessary to calculate the density perturbation and compare the results with the theory of the potential of self-gravitating disk systems (DSS). It should be noted that the nonlinear non-stationarity of model (1) makes it much more difficult to analyze its stability than the corresponding equilibrium disk, since greatly complicates the derivation of NADE in the general case. That is why it is expedient to study the most interesting disturbance modes separately.

## 3. NADE DERIVATION FOR TWO-RING-LIKE OSCILLATION MODES AND THEIR ANALYSIS

3.1. *Case m=0; N=6.* The instability of this mode leads to the formation of two rings in the disk. The following disturbance potential corresponds to this mode:

$$\delta\Phi = D_{06}(\psi)\left(x^2+y^2\right)^3 , \tag{11}$$

and using formula (8), we obtain the components of the centroid displacement in the form

$$\overline{\delta\mathbf{x}} = 6\int_{-\infty}^{\psi} \Pi^3(\psi_1)\mathbf{S}(\psi,\psi_1)\mathbf{D}_{06}(\psi_1)\overline{\mathbf{x}_1\left(\mathbf{x}_1^2+\mathbf{y}_1^2\right)^2}\,\mathbf{d}\psi_1 , \tag{12}$$

$$\overline{\delta\mathbf{y}} = 6\int_{-\infty}^{\psi} \Pi^3(\psi_1)\mathbf{S}(\psi,\psi_1)\mathbf{D}_{06}(\psi_1)\overline{\mathbf{y}_1\left(\mathbf{x}_1^2+\mathbf{y}_1^2\right)^2}\,\mathbf{d}\psi_1 . \tag{13}$$

By definition [1,2,6]

$$\overline{x_1} = xH_\alpha + \bar{u}H_\beta, \qquad \overline{y_1} = yH_\alpha + \bar{\vartheta}H_\beta , \tag{14}$$

where u and $\vartheta$ - are the velocity components in x and y directions, respectively, and

$$H_\alpha = \frac{(\lambda + \cos\psi_1)\cos\psi + \sin\psi \cdot \sin\psi_1}{1 + \lambda\cos\psi}\,,$$
$$H_\beta = \left(1-\lambda^2\right)^{-3/2}[(\lambda + \cos\psi)\sin\psi_1 - (\lambda + \cos\psi_1)\sin\psi]\,. \qquad (15)$$

Then, in accordance with (14), we have

$$\overline{x_1\left(x_1^2+y_1^2\right)^2} = x\left(x^2+y^2\right)^2 H_\alpha^5 + \left[4x\left(x^2+y^2\right)\left(x\bar{u}+y\bar{\vartheta}\right)+\left(x^2+y^2\right)^2\bar{u}\right]H_\alpha^4 H_\beta +$$
$$+\left[2x\left(x^2+y^2\right)\left(\overline{u^2}+\overline{\vartheta^2}\right)+4x\left(x^2\overline{u^2}+2xy\overline{u\vartheta}+y^2\overline{\vartheta^2}\right)+4\left(x^2+y^2\right)\left(x\overline{u^2}+y\overline{u\vartheta}\right)\right]H_\alpha^3 H_\beta^2 +$$
$$+\left[4x\left(x\overline{u^3}+x\overline{u\vartheta^2}+y\overline{u^2\vartheta}+y\overline{\vartheta^3}\right)+2\left(x^2+y^2\right)\left(\overline{u^3}+\overline{u\vartheta^2}\right)+4\left(x^2\overline{u^3}+2xy\overline{u^2\vartheta}+y^2\overline{u\vartheta^2}\right)\right]\cdot$$
$$\cdot H_\alpha^2 H_\beta^3 + \left[x\left(\overline{u^4}+2\overline{u^2\vartheta^2}+\overline{\vartheta^4}\right)+4\left(x\overline{u^4}+x\overline{u^2\vartheta^2}+y\overline{u^3\vartheta}+y\overline{u\vartheta^3}\right)\right]H_\alpha H_\beta^4 +$$
$$+\left[\overline{u^5}+2\overline{u^3\vartheta^2}+\overline{u\vartheta^4}\right]H_\beta^5$$

$$\overline{y_1\left(x_1^2+y_1^2\right)^2} = y\left(x^2+y^2\right)^2 H_\alpha^5 + \left[4y\left(x^2+y^2\right)\left(x\bar{u}+y\bar{\vartheta}\right)+\left(x^2+y^2\right)^2\bar{\vartheta}\right]H_\alpha^4 H_\beta +$$
$$+\left[2y\left(x^2+y^2\right)\left(\overline{u^2}+\overline{\vartheta^2}\right)+4y\left(x^2\overline{u^2}+2xy\overline{u\vartheta}+y^2\overline{\vartheta^2}\right)+4\left(x^2+y^2\right)\left(x\overline{u\vartheta}+y\overline{\vartheta^2}\right)\right]H_\alpha^3 H_\beta^2 +$$
$$+\left[4y\left(x\overline{u^3}+x\overline{u\vartheta^2}+y\overline{u^2\vartheta}+y\overline{\vartheta^3}\right)+2\left(x^2+y^2\right)\left(\overline{\vartheta^3}+\overline{u^2\vartheta}\right)+4\left(x^2\overline{u^2\vartheta}+2xy\overline{u\vartheta^2}+y^2\overline{\vartheta^3}\right)\right]\cdot$$
$$\cdot H_\alpha^2 H_\beta^3 + \left[y\left(\overline{u^4}+2\overline{u^2\vartheta^2}+\overline{\vartheta^4}\right)+4\left(x\overline{u^3\vartheta}+x\overline{u\vartheta^3}+y\overline{u^2\vartheta^2}+y\overline{\vartheta^4}\right)\right]H_\alpha H_\beta^4 +$$
$$+\left[\overline{u^4\vartheta}+2\overline{u^2\vartheta^3}+\overline{\vartheta^5}\right]H_\beta^5$$

For example, the averaging over velocity space for the u component is defined as follows

$$\overline{u^k} = \frac{1}{\sigma(r,t)}\iint u^k \Psi \, du \, d\vartheta\,. \qquad (17)$$

Then we calculate the density response

$$\delta\sigma = -\frac{\partial(\sigma\overline{\delta x})}{\partial x} - \frac{\partial(\sigma\overline{\delta y})}{\partial y} \ . \qquad (18)$$

On the other hand, it is known from the theory of the disk potential that the density perturbation

$$\delta\sigma = \sigma_0 \Pi \cdot \xi^{-1} \cdot P_N^m(\xi)\, e^{im\varphi} \qquad (19)$$

corresponds to the following potential perturbation [8,9]

$$\delta\Phi = 2\Pi^2 \cdot \frac{(N+m-1)!!(N-m-1)!!}{(N+m)!!(N-m)!!} \cdot P_N^m(\xi) \cdot e^{im\varphi} . \qquad (20)$$

Now comparing (19) with m = 0; N = 6 to the calculated result in (18) and taking into account the expressions for (11) and (20), as well as passing from the integral form to the differential form (as proposed in [1]), we obtain the following NADE for this mode

$$\Lambda\ell_\tau(\psi) = \frac{525}{64} D_{06}(\psi) \cdot (\lambda + \cos\psi)^{5-\tau} \sin^\tau\psi, \quad \left(\tau = \overline{0-5}\right) \qquad (21)$$

where the function $D_{06}(\psi)$ is given in Appendix. The resulting NADE (21) is a system of six second-order differential equations. It does not lend itself to analytical consideration and, therefore, was investigated by the method of stability of periodic solutions [10] numerically.

In the course of numerical calculations, changing the values of the rotation parameters Ω, superposition ν and pulsation amplitude λ in the range from 0 to 1, were found the critical values of the initial virial ratio $(2T/|U|)^*_0$, starting from which the model becomes unstable relative to perturbation mode (0,6). The calculation results are presented in the form of dependences of the critical values $(2T/|U|)^*_0$ on Ω and ν (Figure 1).

Numerical analysis of NADE (21) shows that on the background of the composite model, the mode (0; 6) has both oscillatory and aperiodic instabilities for all values of the rotation parameter Ω. The critical diagram of this mode (Figure 1a) for a non-rotating composite model is very similar to the diagram of the single-ring mode (0; 4) [2], and moreover the extreme points A and B lag more behind each other, and in addition, the observed in the case of the mode (0; 4), an additional narrow branch of the instability region, here separately forms a peninsula (with apex at the point S (0.550; 0.588)), not connecting with the main instability zone.

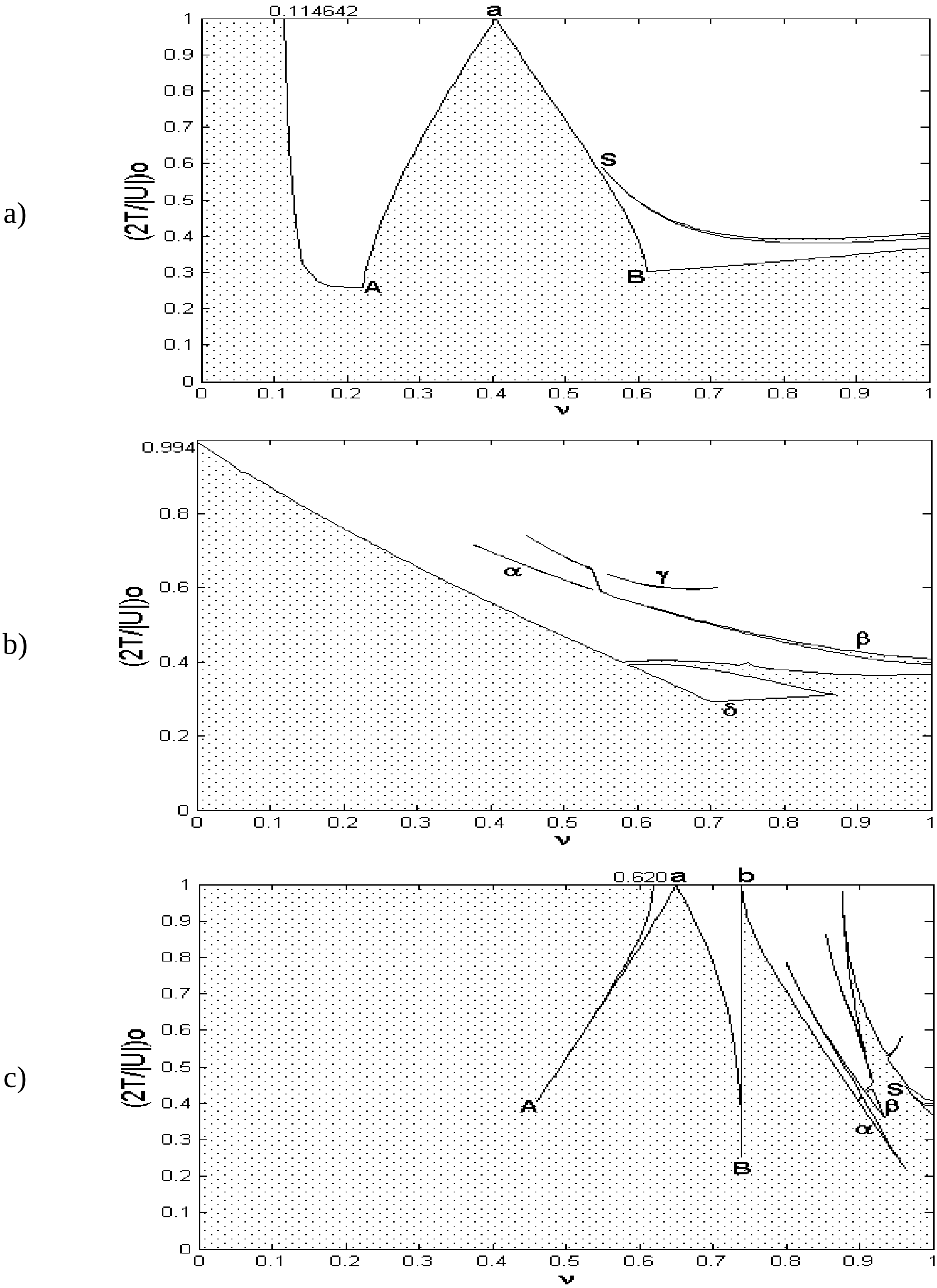

*Fig. 1. Critical dependences of the initial virial ratio on the superposition parameter for the (0;6) mode: a) Ω=0, A(0.220;0.259), B(0.614;0.302), S(0.550;0.588), a=0.404447; b) Ω=0.5; c) Ω=1.0, A(0.46;0.405), B(0.739999;0.254), S1(0.801;0.785), S2(0.855;0.863), S3(0.877103;0.98), a=0.650, b=0.740*

We also note that on the background of a non-rotating composite model, the mode (0; 6) is completely unstable up to the value $\nu \le 0.114642$. Further, the instability region sharply decreases, and when the superposition parameter takes the value $\nu = 0.404447$ this region once again occupies the entire range of possible values of the initial virial ratio. And then, in the interval of $0.405<\nu\le 0.614$, the instability region decreases again to the value $(2T/|U|)_0 \approx 0.302$, and when the superposition parameter tends to its maximum value, it starts to slowly increase again.

The critical diagram at $\Omega=0.5$ has a peculiar form (Figure 1b). In the range of $0.0\le\nu\le 0.58$, the unstable region gradually decreases from $(2T/|U|)_0=0.994$ to $(2T/|U|)_0=0.399$ , and at $0.6 <\nu\le 1$ it is almost stable. In addition, we observe here two islands of instability: $(0.379\le\nu\le 0.54;\ 0.595\le(2T/|U|)_0\le 0.715)$ and $\gamma$ $(0.56\le\nu\le 0.71;\ 0.596\le(2T/|U|)_0\le 0.636$, and one long peninsula - $\beta$ $(0.45\le\nu\le 1.0;\ 0.393\le(2T/|U|)_0\le 0.741)$, as well as one stability island inside the instability region in the form of a "spherical" triangle - $\delta(0.589\le\nu\le 0.87;\ 0.293\le(2T/|U|)_0\le 0.393)$.

At the maximum value of the rotation parameter $\Omega$, the critical diagram (Figure 1c) has two resonant points ($\nu_1=0.650$ and $\nu_2=0.740$) and up to $\nu\le 0.46$ the instability region occupies the entire range of possible values taken by the initial virial ratio. Further, the regions of stability and instability alternate at $(2T/|U|)_0\ge 0.254$. When the superposition parameter approaches its maximum value, stability islands: $\alpha$ $(0.8984\le\nu\le 0.962;\ 0.221\le(2T/|U|)_0\le 0.418)$, $\beta$ $(0.9086\le\nu\le 0.935;\ 0.360\le(2T/|U|)_0\le 0.437)$ and a small peninsula – $\delta(0.963\le\nu\le 1.0;\ 0.368\le(2T/|U|)_0\le 0.447)$ are observed.

*3.2. Case m=2; N=6.* This mode is also responsible for the formation of two rings, but consisting of separate thickenings. In this case, the perturbation potential has the form

$$\delta\Phi = D_{26}(\psi)\left(x^2+y^2\right)^2 (x+iy)^2 . \tag{22}$$

Then the components of the centroid displacement in the perturbed system, according to (8), are defined as

$$\overline{\delta x} = 2\int_{-\infty}^{\psi} \Pi^3(\psi_1) S(\psi,\psi_1) D_{26}(\psi_1)\left[\overline{2x_1\left(x_1^2+y_1^2\right)(x_1+iy_1)^2} + \overline{\left(x_1^2+y_1^2\right)^2 (x_1+iy_1)}\right] d\psi_1$$

$$\overline{\delta y} = 2\int_{-\infty}^{\psi} \Pi^3(\psi_1) S(\psi,\psi_1) D_{26}(\psi_1)\left[\overline{2y_1\left(x_1^2+y_1^2\right)(x_1+iy_1)^2} + i\,\overline{\left(x_1^2+y_1^2\right)^2 (x_1+iy_1)}\right] d\psi_1$$

These expressions for the components of centroid displacement show that averaging is also required for the $\overline{u^3\vartheta}$, $\overline{u^2\vartheta^2}$, $\overline{u\vartheta^3}$, $\overline{u^4}$, $\overline{\vartheta^4}$, $\overline{u^2\vartheta^3}$, $\overline{u^3\vartheta^2}$, $\overline{u\vartheta^4}$, $\overline{u^4\vartheta}$, $\overline{u^5}$, $\overline{\vartheta^5}$. Then, moving to the calculation of the density response and comparing the obtained result with its theoretical expression, we obtain as a result the NADE of the mode (2; 6) against the background of the composite model (1):

$$\Lambda\mu_\tau(\psi)=\frac{105}{256}D_{26}(\psi)\cdot(\lambda+\cos\psi)^{5-\tau}\sin^\tau\psi,\quad \left(\tau=\overline{0-5}\right) \tag{23}$$

Where function of $D_{06}(\psi)$ is given in the Appendix.

The results of numerical calculation of the NADE (23) mode (2; 6) are presented in the form of marginal dependences of the initial virial ratio on the superposition parameter for different values of the rotation parameter in Fig. 2. Using the results of a numerical analysis of NADE (23), we can conclude that the mode (2; 6) by the nature of its instability reminds us of the case of sigle-ring mode (2; 4) whis was investigated by us in [2]. Namely, on the background of a non-rotating composite model (1), there is both an oscillatory and aperiodic instability, but when the model begins to rotate, an instability is observed only with an oscillatory nature.

In the absence of rotation of the composite model (Figure 2a), the mode (2; 6) behaves more stable than the above-considered ring-like modes. There are peninsulas and islands of stability: $\alpha(0\le\nu\le0.0077;\ 0.320\le(2T/|U|)_0\le0.325)$, $\beta(0\le\nu\le0.06;\ 0.292\le(2T/|U|)_0\le0.308)$ and $\delta(0.035\le\nu\le1.0;\ 0.293\le(2T/|U|)_0\le0.473)$,$\gamma(0.5413\le\nu\le0.653;\ 0.227\le(2T/|U|)_0\le0.247)$.

Note that at $\Omega$=0.5 the critical diagram (Figure 2b) of this mode repeats the pattern in the case $\Omega$=0 of mode (2; 4). But only here the peninsula is relatively short and it starts from point S (0.84; 0.552). It also shows a slow increase in the instability region at $\nu$> 0.6, accompanied by an additional branching - S1 (0.66; 0.621). And finally, when the rotation parameter of the composite model takes its maximum value, the critical diagram (Figure 2c) of this mode (2;6) does not qualitatively differ from the considered earlier ring-like modes.

a)

b)

c)

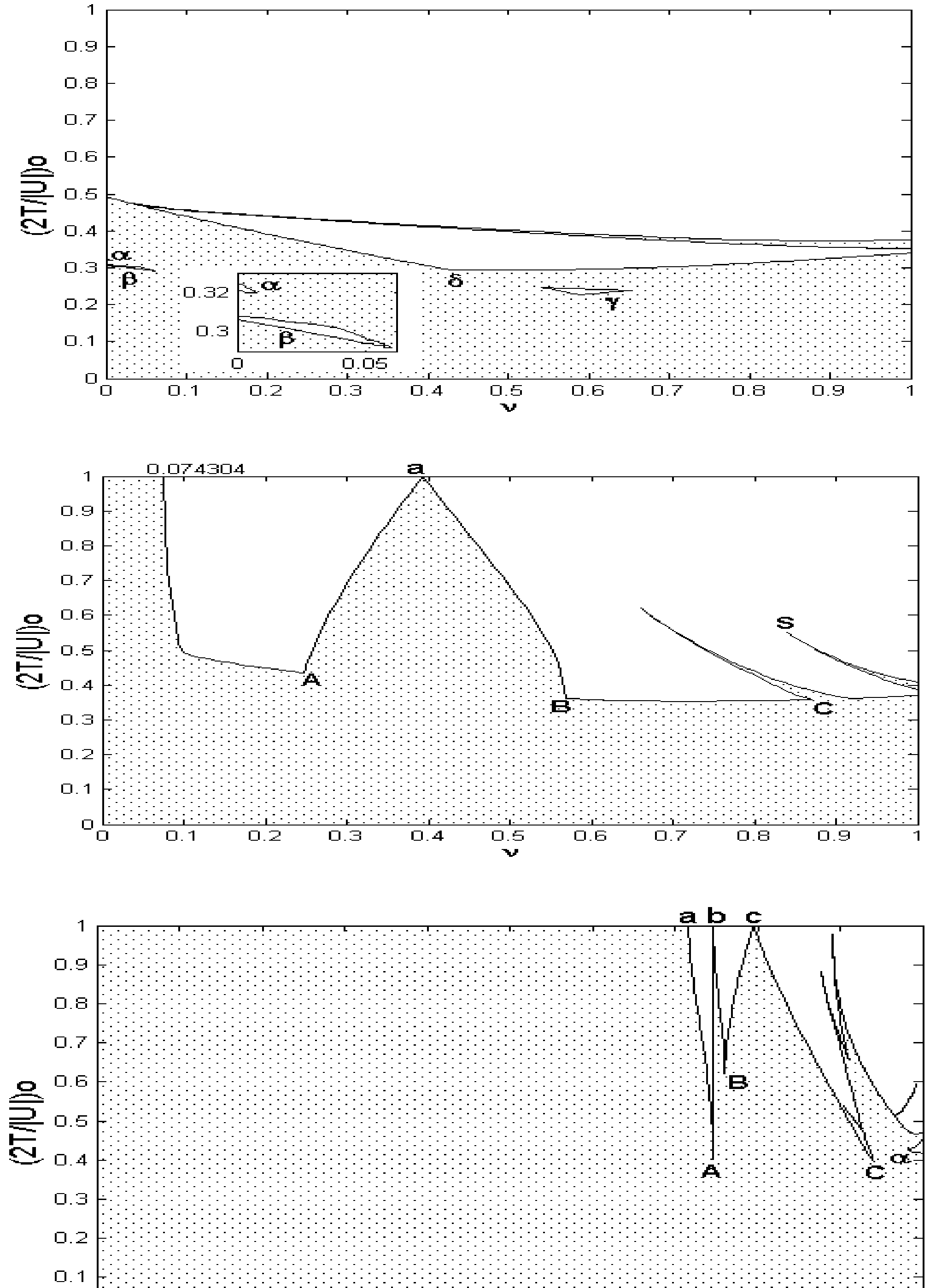

*Fig. 2. Critical dependences of the initial virial ratio on the superposition parameter for the (2;6) mode a) $\Omega$=0, b) $\Omega$=0.5, A(0.2475;0.434), B(0.57;0.361), C(0.87;0.359), a=0.39331, S1(0.66;0.621), S(0.84;0.552); c) $\Omega$=1.0, A(0.74637;0.401), B(0.7588;0.621), C(0.941;0.397), a=0.717051, b=0.746568, c=0.79545, S1(0.877;0.882), S2(0.8902;0.977), $\alpha(0.982 \leq \nu \leq 1; 0.417 \leq (2T/|U|)_0 \leq 0.454)$*

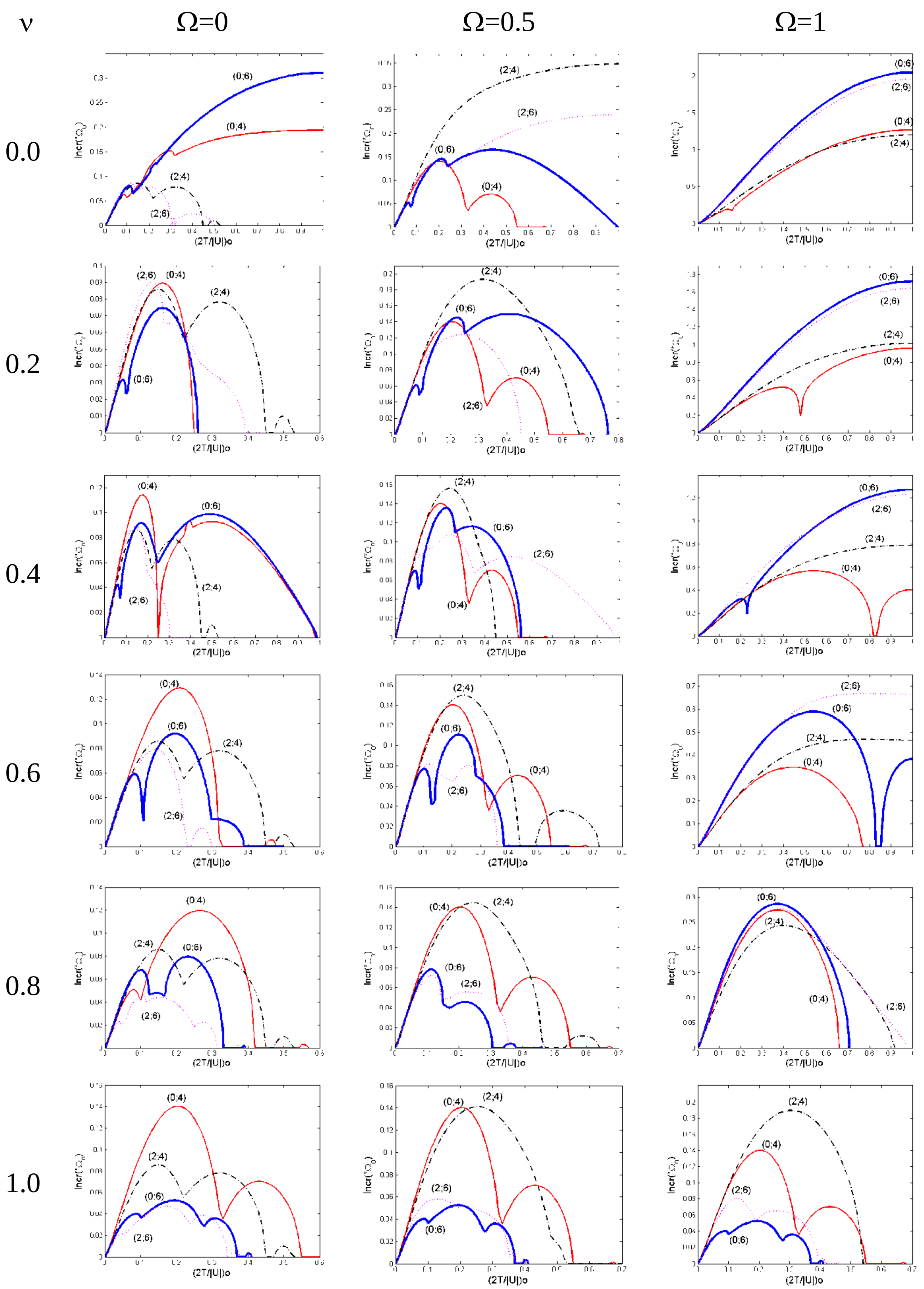

*Fig. 3. Comparision of instability increments for ring-like perturbations modes for different values of the rotation and superposition parameters.*

Figure 3 shows graphs of the comparing of the values of the instability increments of ring-like modes against the background of the composite model (1) for specific values of the rotation parameters Ω and superposition ν. Figure 3 shows that for Ω≠1, single-ring modes are always leading on the background of the composite model, but only in the case of Ω = ν = 0 the two-ring mode (0;6) prevails. Conversely, when the rotation parameter takes its maximum value, the superiority of two-ring modes is observed, but only at ν = 1 single-ring modes becomes leading.

## 4. CONCLUSION

Let us enumerate the main results obtained.

1. We obtained the NADE's for two-ring oscillation modes superimposed on the non-equilibrium model of a self-gravitating disk (1), which is constructed as a superposition of two nonstationary phase densities for DSS with isotropic and anisotropic velocity diagrams.
2. We have determined the marginal dependences between the physical parameters of the model, such as the critical dependences between the values of the initial virial ratio, the degree of rotation, the instability increments and the superposition parameter.
3. It was found that on the background of the composite model, the mode (0; 6) has instabilities of both vibrational and aperiodic nature for all values of the rotation parameter Ω. And in the case of mode (2; 4), on the background of a non-rotating composite model (1), there is both oscillatory and aperiodic instability, but when the model begins to rotate, there is an instability with an oscillatory nature only.
4. It is found that the superposition of the two models leads to a resonance effect, resulting in stretching the instability region up to the value $(2T/|U|)_0 \approx 1$ at the certain values of the superposition parameter.
5. It has been found that as the degree of rotation increases, also increases the range of the initial virial ratio in which the structures under study can form.
6. It is proved that in the non-rotating model (1), the rate of formation of the structure corresponding to the mode (0;4) is greater than in (2;4). And when the model starts to rotate, the picture changes.
7. It is shown that in the absence of rotation of the composite model, the mode (2;6) behaves more stable than the other ring-like modes considered by us. This means that when the system has no rotation, the probability of forming two rings that consist of separate thickenings is very small
8. It turned out that for Ω≠1 on the background of a composite model, single-ring modes are always leading, but only in the case of Ω= ν =0, the two-ring mode (0;6) prevails. Conversely, when the rotation parameter takes its maximum value, the superiority of two-ring modes is observed, but only at ν = 1 are the leading single-ring modes.

**Appendix 1**

$$D_{06}(\psi)=\left[\nu\cdot\left(A_{06}^{*}(\psi)-I_{06}^{*}(\psi)\right)+I_{06}^{*}(\psi)\right]\cdot$$

where

$$I_{06}^{*}(\psi)=\left(g_1\cos^5\psi-g_2\cos^4\psi\sin\psi+g_3\cos^3\psi\sin^2\psi-g_4\cos^2\psi\sin\psi+\right.$$

$$\left.+g_5\cos\psi\sin^4\psi-g_6\sin^5\psi\right)\ell_0(\psi)+\left[5g_1\cos^4\psi\sin\psi+g_2\left(q\cos^4\psi-\right.\right.$$

$$\left.-4\cos^3\psi\sin^2\psi\right)+g_3\left(3\cos^2\psi\sin^3\psi-2q\cos^3\psi\sin\psi\right)+g_4\left(3q\cos^2\psi\sin^2\psi-\right.$$

$$\left.\left.-2\cos\psi\cos^4\psi\right)+g_5\left(\sin^5\psi-4q\cos\psi\sin^3\psi\right)+5g_6q\sin^4\psi\right]\ell_1(\psi)+$$

$$+\left[10g_1\cos^3\psi\sin^2\psi+g_2\left(4q\cos^3\psi\sin\psi-6\cos^2\psi\sin^3\psi\right)+g_3\left(q^2\cos^3\psi-\right.\right.$$

$$-6q\cos^2\psi\sin^2\psi+3\cos\psi\sin^4\psi\Big)+g_4\left(6q\cos\psi\sin^3\psi-3q^2\cos^2\psi\sin\psi-\right.$$

$$\left.\left.-\sin^5\psi\right)+g_5\left(6q^2\cos\psi\sin^2\psi-4q\sin^4\psi\right)-10g_6q^2\sin^3\psi\right]\ell_2(\psi)+$$

$$+\left[10g_1\cos^2\psi\sin^3\psi+g_2\left(6q\cos^2\psi\sin^2\psi-4\cos\psi\sin^4\psi\right)+\right.$$

$$+g_3\left(3q^2\cos^2\psi\sin\psi-6q\cos\psi\sin^3\psi+\sin^5\psi\right)+g_4\left(q^3\cos^2\psi-\right.$$

$$-6q^2\cos\psi\sin^2\psi+3q\sin^4\psi\Big)+g_5\left(6q^2\sin^3\psi-4q^3\cos\psi\sin\psi\right)+$$

$$\left.+10g_6q^3\sin^2\psi\right]\ell_3(\psi)+\left[5g_1\cos\psi\sin^4\psi+g_2\left(4q\cos\psi\sin^3\psi-\sin^5\psi\right)+\right.$$

$$+g_3\left(3q^2\cos\psi\sin^2\psi-2q\sin^4\psi\right)+g_4\left(2q^3\cos\psi\sin\psi-3q^2\sin^3\psi\right)+$$

$$\left.+g_5\left(q^4\cos\psi-4q^3\sin^2\psi\right)-5g_6q^4\sin\psi\right]\ell_4(\psi)+\left(g_1\sin^5\psi+g_2q\sin^4\psi+\right.$$

$$\left.+g_3q^2\sin^3\psi+g_4q^3\sin^2\psi+g_5q^4\sin\psi+g_6q^5\right)\ell_5(\psi),$$

and

$$A_{06}^{*}(\psi)=\frac{h_1^{10}}{8}\left[q\left(8q^4-20e^2q^2\sin^2\psi+5e^4\sin^4\psi\right)\ell_0(\psi)+\right.$$

$$+5e^2\sin\psi\left(e^4\sin^4\psi-16e^2q^2\sin^2\psi+16q^4\right)\ell_1(\psi)+$$

$$+10e^2q\left(23e^2q^2\sin^2\psi-2q^4-8e^4\sin^4\psi\right)\ell_2(\psi)+$$

$$+10e^4\sin\psi\left(23e^2q^2\sin^2\psi-2e^4\sin^4\psi-8q^4\right)\ell_3(\psi)$$

$$+5e^4q\left(q^4-16e^2q^2\sin^2\psi+16e^4\sin^4\psi\right)\ell_4(\psi)$$

$$\left.+e^6\sin\psi\left(8e^4\sin^4\psi-20e^2q^2\sin^2\psi+5q^4\right)\ell_5(\psi)\right].$$

while

$$\mathrm{h}_1=(1+\lambda\cos\psi)^{-1};\qquad \mathrm{c}=\lambda\sin\psi\Big/\sqrt{1-\lambda^2}$$

$$\mathbf{q}=\boldsymbol{\lambda}+\cos\boldsymbol{\psi},\qquad \mathbf{e}=\sqrt{1-\boldsymbol{\lambda}^2}$$

$$\mathrm{g}_1=\mathbf{h}_1^5;\qquad \mathrm{g}_2=-5\mathrm{c}\sqrt{1-\lambda^2}\,\mathrm{h}_1^6;\qquad \mathrm{g}_3=2\left(3\Omega^2+5\mathrm{c}^2-2\right)\left(1-\lambda^2\right)\mathrm{h}_1^7;$$

$$\mathbf{g}_4=2\mathbf{c}\left(6-9\Omega^2-5\mathrm{c}^2\right)\left(1-\lambda^2\right)^{3/2}\mathrm{h}_1^8;$$

$$\mathbf{g}_5=\frac{1}{5}\left(8-36\Omega^2+33\Omega^4-60\mathrm{c}^2+90\mathbf{c}^2\Omega^2+25\mathbf{c}^4\right)\left(1-\lambda^2\right)^2\mathrm{h}_1^9;$$

$$\mathbf{g}_6=\frac{\mathbf{c}}{5}\left(20\mathrm{c}^2-30\mathbf{c}^2\Omega^2-5\mathbf{c}^4-8+36\Omega^2-33\Omega^4\right)\left(1-\lambda^2\right)^{5/2}\mathrm{h}_1^{10}.$$

and

$$\ell_\tau(\psi)=\int_{-\infty}^{\psi}\left(1+\lambda cos\psi_1\right)^3 S(\psi,\psi_1)D_{06}(\psi_1)(\lambda+cos\psi_1)^{5-\tau}sin^\tau\psi_1 d\psi_1$$

$$D_{06}(\psi)=\left[\nu\cdot\left(A_{06}^{*}(\psi)-I_{06}^{*}(\psi)\right)+I_{06}^{*}(\psi)\right]\cdot$$

where

$$I_{26}^{*}(\psi)=\left(b_1\cos^5\psi-b_2\cos^4\psi\sin\psi+b_3\cos^3\psi\sin^2\psi-b_4\cos^2\psi\sin\psi+\right.$$

$$\left.+b_5\cos\psi\sin^4\psi-b_6\sin^5\psi\right)\mu_0(\psi)+\left[5b_1\cos^4\psi\sin\psi+b_2\left(q\cos^4\psi-\right.\right.$$

$$\left.-4\cos^3\psi\sin^2\psi\right)+b_3\left(3\cos^2\psi\sin^3\psi-2q\cos^3\psi\sin\psi\right)+b_4\left(3q\cos^2\psi\sin^2\psi-\right.$$

$$\left.\left.-2\cos\psi\cos^4\psi\right)+b_5\left(\sin^5\psi-4q\cos\psi\sin^3\psi\right)+5b_6q\sin^4\psi\right]\mu_1(\psi)+$$

$$+\left[10b_1\cos^3\psi\sin^2\psi+b_2\left(4q\cos^3\psi\sin\psi-6\cos^2\psi\sin^3\psi\right)+b_3\left(q^2\cos^3\psi-\right.\right.$$

$$\left.-6q\cos^2\psi\sin^2\psi+3\cos\psi\sin^4\psi\right)+b_4\left(6q\cos\psi\sin^3\psi-3q^2\cos^2\psi\sin\psi-\right.$$

$$\left.\left.-\sin^5\psi\right)+b_5\left(6q^2\cos\psi\sin^2\psi-4q\sin^4\psi\right)-10b_6q^2\sin^3\psi\right]\mu_2(\psi)+$$

$$+\left[10b_1\cos^2\psi\sin^3\psi+b_2\left(6q\cos^2\psi\sin^2\psi-4\cos\psi\sin^4\psi\right)+\right.$$

$$+b_3\left(3q^2\cos^2\psi\sin\psi-6q\cos\psi\sin^3\psi+\sin^5\psi\right)+b_4\left(q^3\cos^2\psi-\right.$$

$$\left.-6q^2\cos\psi\sin^2\psi+3q\sin^4\psi\right)+b_5\left(6q^2\sin^3\psi-4q^3\cos\psi\sin\psi\right)+$$

$$\left.+10b_6q^3\sin^2\psi\right]\mu_3(\psi)+\left[5b_1\cos\psi\sin^4\psi+b_2\left(4q\cos\psi\sin^3\psi-\sin^5\psi\right)+\right.$$

$$+b_3\left(3q^2\cos\psi\sin^2\psi-2q\sin^4\psi\right)+b_4\left(2q^3\cos\psi\sin\psi-3q^2\sin^3\psi\right)+$$

$$\left.+b_5\left(q^4\cos\psi-4q^3\sin^2\psi\right)-5b_6q^4\sin\psi\right]\mu_4(\psi)+\left(b_1\sin^5\psi+b_2q\sin^4\psi+\right.$$

$$\left.+b_3q^2\sin^3\psi+b_4q^3\sin^2\psi+b_5q^4\sin\psi+b_6q^5\right)\mu_5(\psi),$$

and

$$A_{26}^{*}(\psi)=\frac{h_1^{10}}{64}\left\{\left[152q\left(8q^4-20e^2q^2\sin^2\psi+5e^4\sin^4\psi\right)+\right.\right.$$

$$\left.+37i\Omega e\sin\psi\left(16e^2q^2\sin^2\psi-16q^4-e^4\sin^4\psi\right)\right]\mu_0(\psi)+$$

$$+\left[760e^2\sin\psi\left(e^4\sin^4\psi-16e^2q^2\sin^2\psi+16q^4\right)+\right.$$

$$+37i\Omega eq\left(16q^4+37e^4\sin^4\psi-112e^2q^2\sin^2\psi\right)\Big]\mu_1(\psi)+$$

$$+\Big[1520e^2q\left(23e^2q^2\sin^2\psi-2q^4-8e^4\sin^4\psi\right)+$$

$$+74i\Omega e^3\sin\psi\left(56q^4-101e^2q^2\sin^2\psi+8e^4\sin^4\psi\right)\Big]\mu_2(\psi)+$$

$$+\Big[1520e^4\sin\psi\left(23e^2\sin^2\psi-2e^4\sin^4\psi-8q^4\right)+$$

$$+74i\Omega e^3q\left(101e^2q^2\sin^2\psi-56e^4\sin^4\psi-8q^4\right)\Big]\mu_3(\psi)+$$

$$+\Big[760e^4q\left(q^4-16e^2q^2\sin^2\psi+16e^4\sin^4\psi\right)+$$

$$+37i\Omega e^5\sin\psi\left(112e^2q^2\sin^2\psi-16e^4\sin^4\psi-37q^4\right)\Big]\mu_4(\psi)+$$

$$+\Big[152e^6\sin\psi\left(8e^4\sin^4\psi-20e^2q^2\sin^2\psi+5q^4\right)+$$

$$+37i\Omega e^5q\left(16e^4\sin^4\psi-16e^2q^2\sin^2\psi+q^4\right)\Big]\mu_5(\psi)\Big\}.$$

$$b_1=19\mathbf{h}_1^5;\qquad b_2=(-95c+37i\Omega)\sqrt{1-\lambda^2}\,h_1^6;$$

$$b_3=\left[2\left(33\Omega^2+95c^2-32\right)-148\mathbf{i\Omega c}\right]\left(1-\lambda^2\right)h_1^7;$$

$$\mathbf{b}_4=2\left[\mathbf{c}\left(96-99\Omega^2-95c^2\right)+\mathbf{i}\Omega\left(111\mathbf{c}^2+75\Omega^2-56\right)\right]\left(1-\lambda^2\right)^{3/2}h_1^8;$$

$$\mathbf{b}_5=\left[16-33\Omega^4-192c^2+198\mathbf{c}^2\Omega^2+95\mathbf{c}^4+4\mathbf{i\Omega c}\left(37\mathbf{c}^2-75\Omega^2+56\right)\right]\times$$

$$\times\left(1-\lambda^2\right)^2h_1^9,$$

$$\mathbf{b}_6=\left[\mathbf{c}\left(64c^2-66\mathbf{c}^2\Omega^2-19\mathbf{c}^4-16+33\Omega^4\right)+\right.$$

$$\left.+\mathbf{i}\Omega\left(37\mathbf{c}^4+150\mathbf{c}^2\Omega^2-112\mathbf{c}^2+16-48\Omega^2+33\Omega^4\right)\right]\left(1-\lambda^2\right)^{5/2}h_1^{10}.$$

$$\mu_\tau(\psi)=\int_{-\infty}^{\psi}\left(1+\lambda\cos\psi_1\right)^3S(\psi,\psi_1)D_{26}(\psi_1)\left(\lambda+\cos\psi_1\right)^{5-\tau}\sin^\tau\psi_1\,d\psi_1$$